\documentclass[a4paper,fleqn,usenatbib]{mnras}

\usepackage[T1]{fontenc}
\usepackage{ae,aecompl}

\usepackage{graphicx}	
\usepackage{amsmath}	
\usepackage{amssymb}	

\title[The Host Galaxy of HESS~J1943+213]{
Resolving the Host Galaxy of a Distant Blazar with LBT/LUCI\,1 + ARGOS}

\author[E.\ P.\ Farina et al.]{\parbox{\textwidth}{
E.~P.~Farina,$^{1}$\thanks{E-mail: emanuele.paolo.farina@gmail.com}
I.~Y.~Georgiev,$^{1}$
R.~Decarli,$^{1,2}$
T.~Terzi{\'c},$^{3}$
L.~Busoni,$^{4}$
W.~G{\"a}ssler,$^{1}$
T.~Mazzoni,$^{4}$
J.~Borelli,$^{1}$
M.~Rosensteiner,$^{5}$
J.~Ziegleder,$^{5}$
M.~Bonaglia,$^{4}$
S.~Rabien,$^{5}$
P.~Buschkamp,$^{6}$
G.~Orban~de~Xivry,$^{5}$
G.~Rahmer,$^{7}$
M.~Kulas,$^{1}$ and
D.~Peter.$^{1,8}$}
\vspace{0.4cm}\\
$^{1}$Max Planck Institut f{\"u}r Astronomie, K{\"o}nigstuhl 17, D--69117, Heidelberg, Germany\\
$^{2}$Osservatorio Astronomico di Bologna, via Gobetti 93/3, I--40129, Bologna, Italy\\
$^{3}$Department of Physics, University of Rijeka, Radmile Matej\v{c}i\'{c} 2, HR--51000, Rijeka, Croatia\\
$^{4}$Osservatorio Astrofisico di Arcetri, Largo Enrico Fermi 5, I--50125, Florence, Italy\\
$^{5}$Max Planck Institut f{\"u}r Extraterrestrische Physik, Giessenbachstrasse 1, D--85748, Garching, Germany\\
$^{6}$Buschkamp Research Instruments, Elisabethstrasse 2, D--80796, M{\"u}nchen, Germany\\
$^{7}$LBT Observatory, University of Arizona, 933 N.\ Cherry Ave, AZ--85721, Tucson, USA\\
$^{8}$Heidelberg Instruments Mikrotechnik GmbH, Tullastrasse 2, D--69126, Heidelberg, Germany
}

\date{Accepted XXX. Received YYY; in original form ZZZ}

\pubyear{2017}

\begin{document}
\label{firstpage}
\pagerange{\pageref{firstpage}--\pageref{lastpage}}
\maketitle

\begin{abstract}
BL Lac objects emitting in the Very High Energy (VHE) regime are unique tools to peer into the properties of the Extragalactic Background Light (EBL).
However, due to the typical absence of features in their spectra, the determination of their redshifts has proven challenging.
In this work we exploit the superb spatial resolution delivered by the new Advanced Rayleigh guided Ground layer adaptive Optics System (\textit{ARGOS}) at the Large Binocular Telescope to detect the host galaxy of HESS~J1943+213, a VHE emitting BL Lac shining through the Galaxy.
Deep $H$--band imaging collected during the ARGOS commissioning allowed us to separate the contribution of the nuclear emission and to unveil the properties of the host galaxy with unprecedented detail.
The host galaxy is well fitted by a S{\'e}rsic profile with index of $n$$\sim$2 and total magnitude of $H$$_{\rm Host}$$\sim$16.15\,mag.
Under the assumption that BL~Lac host galaxies are standard candles, we infer a redshift of $z$$\sim$0.21.
In the framework of the current model for the EBL, this value is in agreement with the observed dimming of the VHE spectrum due to the scatter of energetic photons on the EBL.
\end{abstract}

\begin{keywords}
instrumentation: adaptive optics -- infrared: galaxies -- BL Lacertae objects: individual: HESS~J1943+213
\end{keywords}


\section{Introduction}

In the classical unified model, blazars constitute a class of active galactic nuclei (AGN) viewed at small angles from the jet axis
\citep[][]{Blandford1978, Antonucci1993, Urry1995}.
Traditionally, blazars has been further splitted into two sub--classes based on the strength of the features present in their optical spectra.
While flat--spectrum radio quasars (FSRQs) show emission lines with equivalent width $\gtrsim$$5$\,\AA, in BL Lacertae objects (BL Lacs) the non--thermal synchrotron radiation of the jet completely dominates the optical/UV emission, ending up in a typical featureless power--law spectrum.
This makes the determination of their redshift via the detection of absorption/emission lines from the nuclear emission and/or from the host galaxy particularly challenging, even with 8--10 meters class telescopes (e.g. \citealt{Sbarufatti2005a, Sbarufatti2006, Sbarufatti2009, Landoni2013, Sandrinelli2013, Shaw2013, Pita2014, RosaGonzalez2016, Paiano2016}; and \citealt{Falomo2014} for a review).
In past years, several alternatives have been proposed to constrain the redshift of BL Lac objects, including: 
the detection of intervening absorption features either from the halo of lower redshift galaxies \citep[e.g.][]{Shaw2013, Landoni2014} or from the neutral hydrogen in the intergalactic medium \citep[e.g.][]{Danforth2010, Furniss2013};
the spectroscopy of galaxies in the environment where the blazars are embedded \citep[e.g.][]{Muriel2015, Farina2016};
the detection of molecular emission lines from the host galaxy \citep[e.g.][]{Fumagalli2012};
the study of the effect of the interaction with the extragalactic background light (EBL) in the blazar emission in the GeV and TeV domain \citep[e.g.][]{Prandini2010, Prandini2012}.
In particular, the narrow distribution in luminosity of BL Lac host galaxies \citep[][]{Urry2000, Sbarufatti2005b} opened the possibility to use them as standard candles, and thus to measure their distance via broad band imaging \citep[e.g.][]{Nilsson2008, Meisner2010, Kotilainen2011}.
The main challenge of this approach is to accurately remove the bright central emission that typically outshine the host galaxy.
Given the average 1\farcs0 (or 3.2\,kpc in the considered cosmology) effective radius calculated from the collection of $z$$\lesssim$0.6 BL~Lac hosts observed with {\it HST} by \citet{Scarpa2000}, it is clear that images with an exquisite spatial resolution and high contrast are necessary to unveil the faint and diffuse starlight emission around the bright, point--like emission from the active nucleus.
In this paper we exploit the capabilities of the new Advanced Rayleigh guided Ground layer adaptive Optics System \citep[{\it ARGOS};][]{Rabien2010} mounted on the Large Binocular Telescope \citep[LBT,][]{Hill2004, Hill2012} to collect high--resolution near--infrared (NIR) LUCI\,1 \citep[i.e., LBT Utility Camera in the Infrared;][]{Seifert2003, Ageorges2010} observations of HESS~J1943+213.
This blazar was detected by {\it H.E.S.S.} in the very high--energy domain (VHE, i.e. at $E$$>$100\,GeV) during a VHE galactic survey \citep[][]{HESS2011}, making it the only BL~Lac object known located in the Galactic plane.
Broad $Ks$--band images gathered with the 3.5\,m CAHA telescope revealed the presence of an extended emission that has been attributed to the host galaxy of HESS~J1943+21 \citep{Peter2014}.
A comparison with the typical size of blazar host derived by \citet{Cheung2003} allowed \citet{Peter2014} to set a lower limit on the redshift of $z$$>$0.03.
This is consistent with the $z$$>$0.14 derived from the fit of the spectral energy distribution of HESS~J1943+21 \citep[][]{HESS2011, Cerruti2011} and with the $z$$<$0.45 limit obtained via modelling the attenuation of the VHE emission by the EBL \citep{Peter2014}.
A tighter constraint on the redshift is however necessary to understand the nature of the VHE emission and to derive the EBL properties.

Throughout this paper, we assume a concordance cosmology with H$_0$=70\,km\,s$^{-1}$\,Mpc$^{-1}$, $\Omega_{\rm m}$=0.3, and $\Omega_\Lambda$=0.7.
All the quoted magnitudes are expressed in the AB standard photometric system \citep[][]{Oke1974, Oke1983}.

\section{Observations and Data Reduction}\label{sec:dr}

\begin{figure}
\begin{center}
\includegraphics[width=0.49\textwidth]{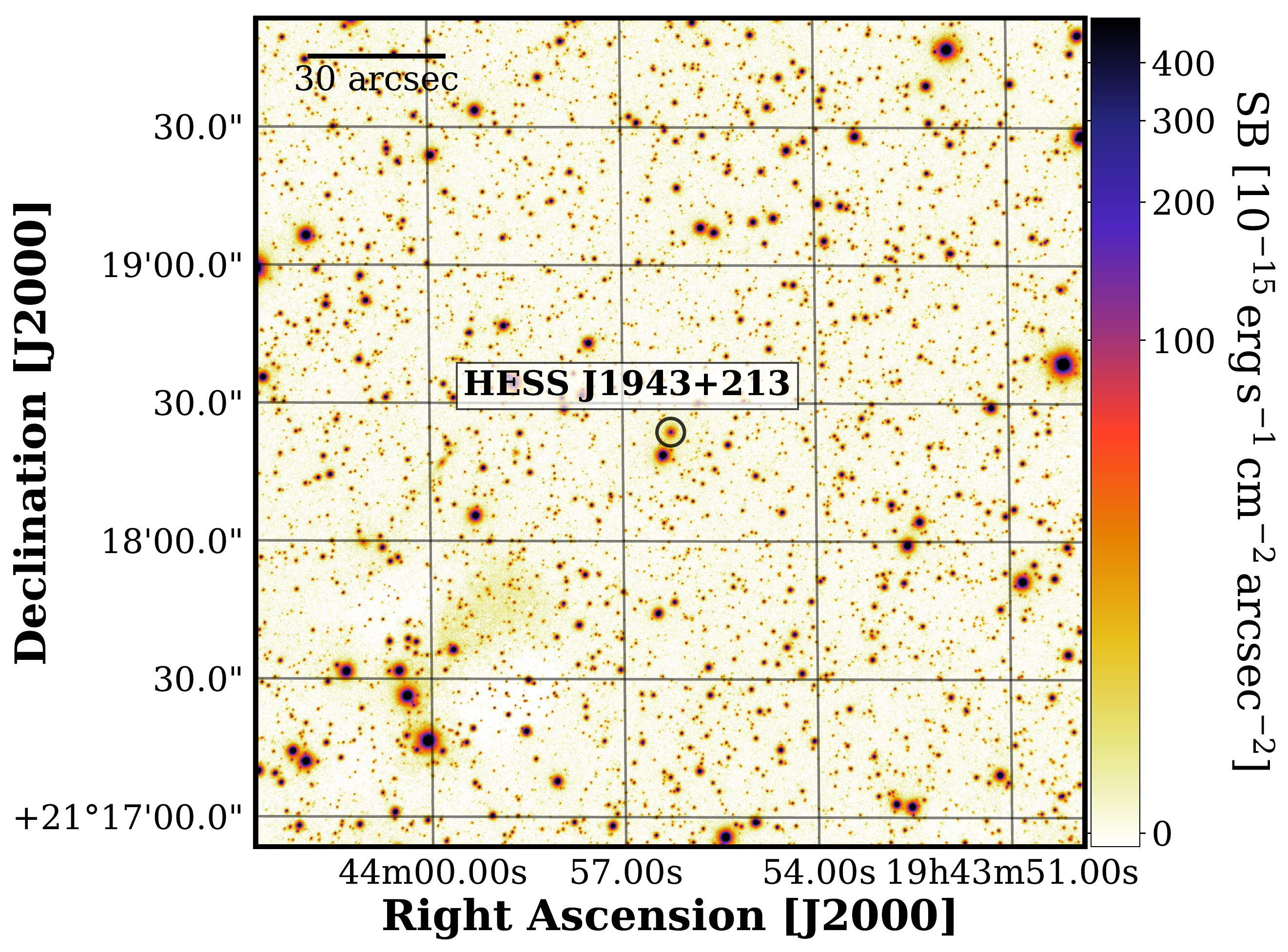}
\caption{
The central 3$^\prime$$\times$3$^\prime$ region around HESS~J1943+213 as imaged in the $H$--Band with LBT/LUCI\,1+ARGOS (North is up and West is to the right).
Marked with a circle is the position of the blazar  \citep[corresponding to the source located at R.A.$_{\rm J2000}$=19:43:57.63 and Dec.$_{\rm J2000}$=+21:18:31.5 in the 2MASS catalogue][]{Cutri2003}.
The large field of view of LBT/LUCI\,1+ARGOS and the location of the source close to the Galactic plane (with Galactic latitude $b$=$-1\fdg 2947$) allowed us to sample well and build good PSF model (see Section~\ref{sec:psf}).
}\label{fig:fov}
\end{center}
\end{figure}

NIR data in the $H$--band with the LBT/LUCI\,1 NIR camera were collected during the ARGOS commissioning run on 2016-10-20.
ARGOS uses a natural guide star (NGS) for guiding, while three green light (532\,nm) lasers focused at 12\,km (the laser guide stars, LGSs) are used to correct for the ground layer turbulence.
The LGSs are situated on a circle with radius 2\arcmin\ which provides AO correction across the entire 4\arcmin$\times$4\arcmin\ LUCI\,1 field of view (see Figure~\ref{fig:fov}).
The NGS that we used is conveniently bright for guiding ($R$=$16.5$\,mag) but too faint for fast wave front sensing without the lasers.
The location of the NGS was close ($\Delta r$=$5\arcsec$) to our target and it was placed in the centre of the LUCI\,1 field.
This provides an excellent condition for the optimal ARGOS AO performance and stability of the PSF (see \S\,\ref{sec:psf}).
Naturally, the seeing in the optical is higher than that in the NIR and during the night it was around 1\farcs0$\pm$0\farcs1, while that in the $H$--band was around 0\farcs45$\pm$0\farcs05. 
The use of ARGOS halved the natural seeing in the $H$-band to 0\farcs26 (FWHM=2.2\,pixel). 
To obtain good sky subtraction, we used a 30\arcsec$\times$30\arcsec dither box.
This is significantly larger ($>$5$\times$) than the angular size of our target, and accounts also for the nearby bright foreground star.
Each offset and exposure acquisition was completed to within one minute, which allowed to combine five offsets and map well the temporal variation of the NIR night sky.
Data reduction was performed within the {\textsc IRAF} environment.
Master dark and flat field calibration data were used to reduce the raw images.
Correction for persistence and non--linearity was performed using pixel maps as described in \citet{Georgiev2017}.
In total, we combined 228 individual exposures, each of DIT=2.8 seconds.
The short DIT was required to avoid badly saturating foreground stars and avoiding strong persistence.
The final image registration and combination was performed with our custom wrapper routine built around the main {\textsc IRAF} tasks {\sc geomap, geotran, imcombine}.
During the final geometric image transformation and combination we used a drizzle drop fraction of 0.8 and a 20$\times$20 pixels statistics region to check additionally for residual background variation between the individual exposures.

The resulting image\footnote{The reduced image is publicly available at: \href{https://github.com/EmAstro/LBT_ARGOS}{\texttt{https://github. com/EmAstro/LBT\_ARGOS}}.} (see Figure~\ref{fig:fov}) was then WCS registered using the {\textsc Astrometry.net} tool \citep[][]{Lang2010} and calibrated in flux matching sources detected in the field with the 2MASS catalogue \citep[][]{Cutri2003}.
We adopted the conversion from Vega to AB magnitudes from \citet[][]{Blanton2005}.
Uncertainties in the zeropoint are of the order of 0.06\,mag.
The 5--$\sigma$ detection limit for a point source (estimated from the rms of the sky counts integrated over the radius of an unresolved source, i.e. 1.1\,pixel) is $H$$_{\rm lim}$$\approx$23.5\,mag.
We consider the reddening correction from \citet{Schlafly2011} [$E(B-V)$=2.30\,mag, toward the location of the blazar].
Assuming a visual extinction to reddening ratio of $R_{V}$=$A_{V}$/$E(B-V)$=3.1 \citep[e.g.][]{Cardelli1989, Fitzpatrick1999} and the \citet{Cardelli1989} extinction curve, this corresponds to a $H$--band extinction of $A^{c}_{H}$=1.30\,mag. 
The dust extinction in this region of the Galaxy, however, is not well constrained.
For instance, a much lower extinction ($E(B-V)$=1.76\,mag) is reported by \citealt[][]{Green2015} for the same region of the sky.
In the following, we thus consider that the $H$--band extinction can vary from $A^{c}_{H,{\rm min}}$=1.19\,mag and $A^{c}_{H,{\rm Max}}$=1.62\,mag, corresponding respectively to the minimum and maximum value of $E(B-V)$ observed within a circle of 5\,arcmin radius centred at the location of HESS~J1943+213 \citep[][]{Schlafly2011}.

\section{Subtraction of the PSF}\label{sec:psf}

\begin{figure*}
\begin{center}
\includegraphics[width=1.1\textwidth, bb=1.5in 0.6in 21.0in 8.0in]{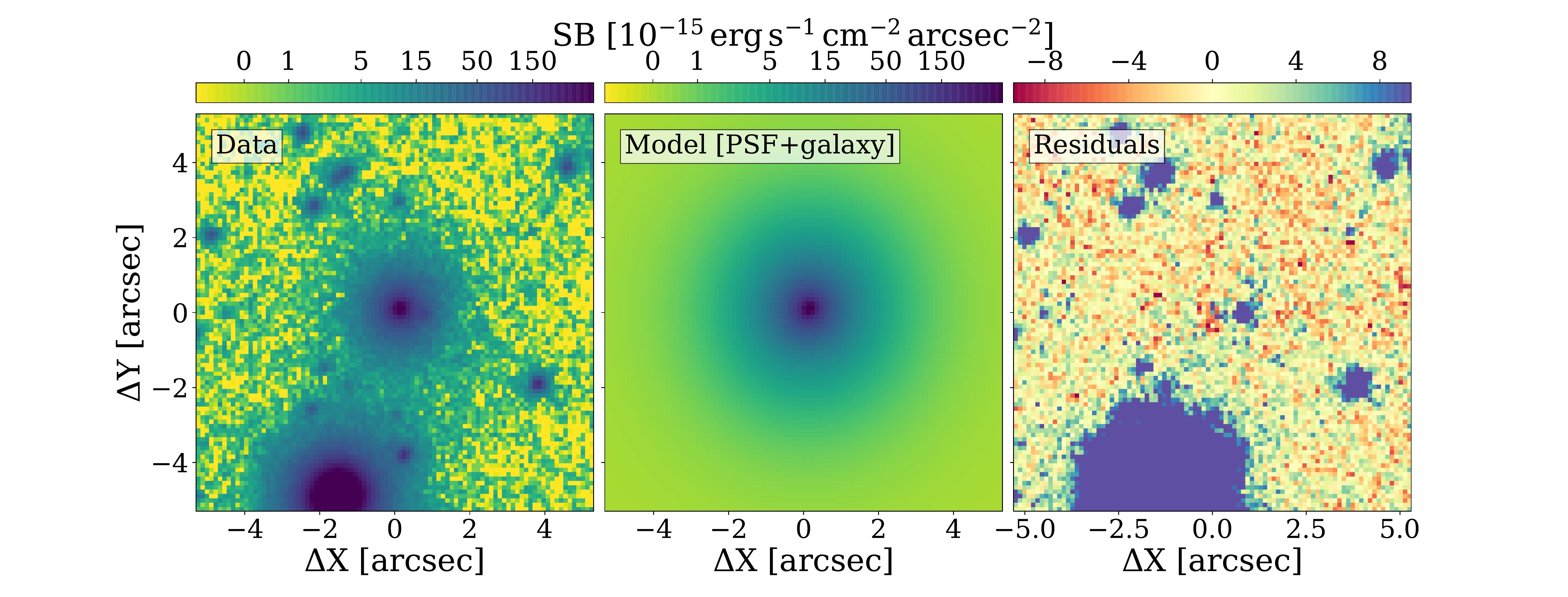}
\caption{
Results from the point source and host galaxy modelling of the blazar HESS~J1943+213.
{\it Left --- } Thumbnail of the LBT/LUCI\,1+ARGOS $H$--band image centred at the blazar location. 
For display purposes, the best fit of the background provided by {\textsc GALFIT} has been removed.
In all panels, North is up and West is right. 
{\it Central --- } The best fit {\textsc GALFIT} model.
The PSF model was created as described in Section~\ref{sec:psf}.
The host galaxy light appears to follow a S{\'e}rsic profile with index $n$=2.2 and effective radius $R_e$$\sim$1.1\,arcsec.
The derived total magnitude of the host galaxy is: H$_{\rm Host}$=16.15\,mag (not corrected for Galactic extinction).
An independent analysis performed with {\textsc Imfit} results in a similar solution for the fit. 
{\it Right --- } Residuals after model subtraction.
The residual image shows a source located at 0\farcs6 West from the centre of the blazar. We masked it during the fitting process in order to avoid contaminating the extended emission from the host galaxy.
}
\label{fig:res}
\end{center}
\end{figure*}

\begin{figure}
\begin{center}
\includegraphics[width=0.5\textwidth, bb=0.45in 0.65in 8.0in 8.0in]{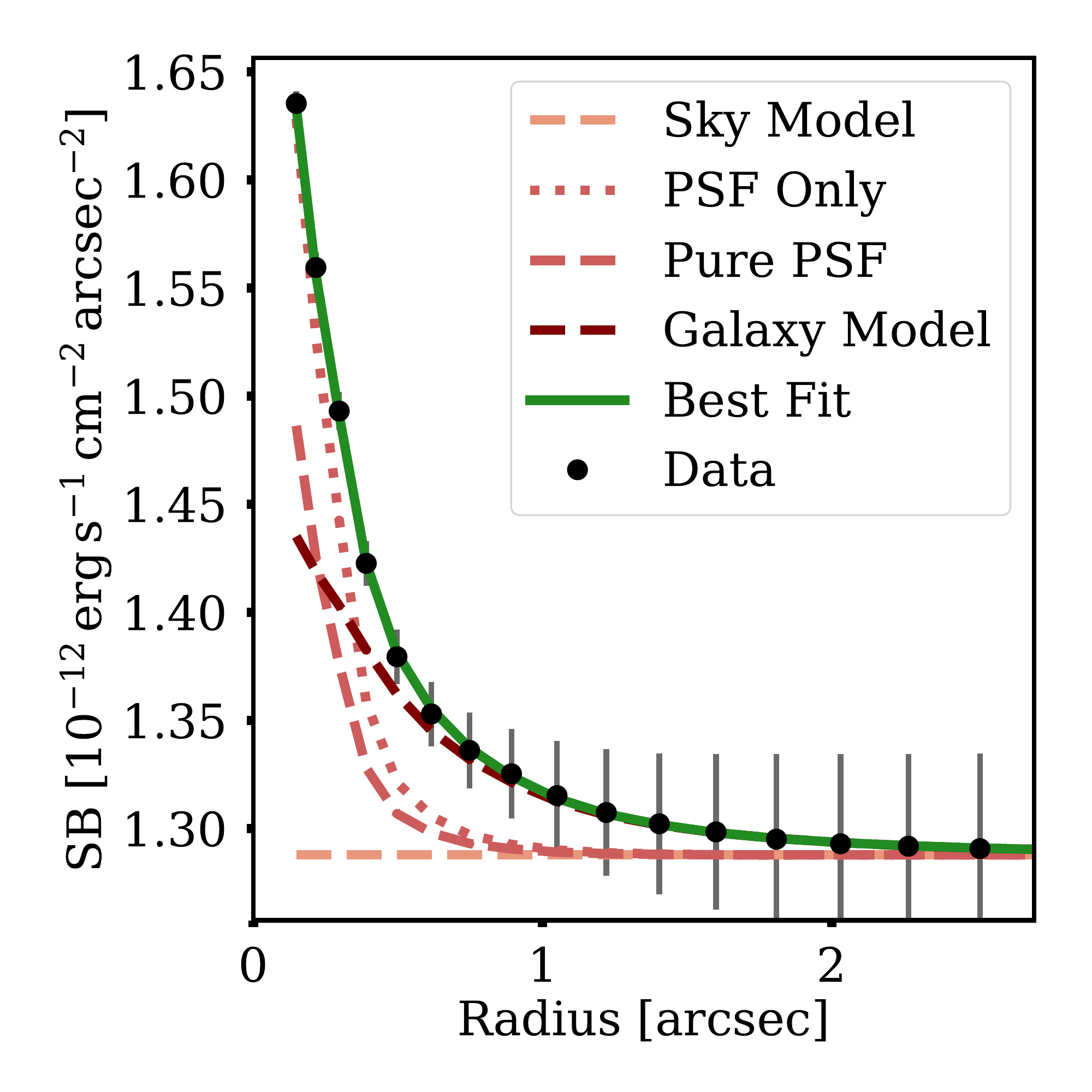}
\caption
{
Radial intensity profile of the BL Lac emission extracted in circular annulii (black points).
The different components used to model the emission of HESS~J1943+213 are marked with different coloured dashed lines (see Figure~\ref{fig:res} and Section~\ref{sec:psf}).
The total fit is shown as a green solid line.
The dotted line illustrates the result of the fitting procedure assuming that no extended emission is present in the data.
It is evident that the host galaxy emission is much more extended than the PSF.
}
\label{fig:sb}
\end{center}
\end{figure}

\begin{table}
\centering
\caption{Result from the {\textsc GALFIT} fit of the blazar HESS~J1943+213.
Magnitudes are not corrected for Galaxy extinction.}
\label{tab:fit}
\begin{tabular}{ll} 
	\hline
	Parameter        & Value        \\
	\hline
	$H$$_{\rm PSF}$  & $\left(18.10^{+0.07}_{-0.05}\right)$\,mag  \\
	$H$$_{\rm Host}$ & $\left(16.15^{+0.01}_{-0.03}\right)$\,mag  \\
	$R_e$            & $\left(1.13^{+0.02}_{-0.04}\right)$\,arcsec \\
	S{\'e}rsic index & $\left(2.20^{+0.31}_{-0.23}\right)$         \\
	Axis Ratio       & $\left(0.92^{+0.01}_{-0.01}\right)$         \\
	Position Angle   & $\left(-1\fdg49^{+0\d\fdg12}_{-3\fdg73}\right)$  \\
	\hline
\end{tabular}
\end{table}

The model of the Point Spread Function (PSF) was built by using 68 Milky Way foreground stars. They were chosen to have no contamination from neighbours within the radius of the PSF model, to sample well the detector and build a spatially variable PSF. As mentioned in Section\,\ref{sec:dr}, the NGS is projected very close to our blazar and is in the centre of the LUCI\,1 field. This guarantees a very stable and sharp PSF, although, as shown in \cite{Georgiev2017}, its global variation is up to 25\% toward the detector edges.

This PSF model was then ingested into {\textsc GALFIT} \citep[version 3.0.5,][]{Peng2010} in order to infer the properties of the extended emission around HESS~J1943+213.
To avoid contamination from nearby objects, we masked all sources present within a 10\arcsec\ radius from the blazar, including the faint, unresolved object located 0\farcs6 West from HESS~J1943+213 (see Figure~\ref{fig:res}).
As a first step, we assumed the source as unresolved.
In this case the entire emission would fall within our PSF model.
The fit of a pure PSF (representing the central, unresolved nuclear emission) and of the sky background leave, however, significant residuals, confirming the presence of an extended emission (see Figure~\ref{fig:sb}).
In addition to the PSF model and to the sky emission we thus simultaneously fitted a galaxy component modelled with a S{\'e}rsic profile \citep[][]{Sersic1963}.
This second approach leave only negligible residuals.
Results of the fitting procedure are showed in Figures~\ref{fig:res} and~\ref{fig:sb} and summarised in Table~\ref{tab:fit}.

\section{Properties of the Host Galaxy}\label{sec:host}

\begin{figure}
\begin{center}
\includegraphics[width=0.5\textwidth, bb=0.45in 0.65in 8.0in 8.0in]{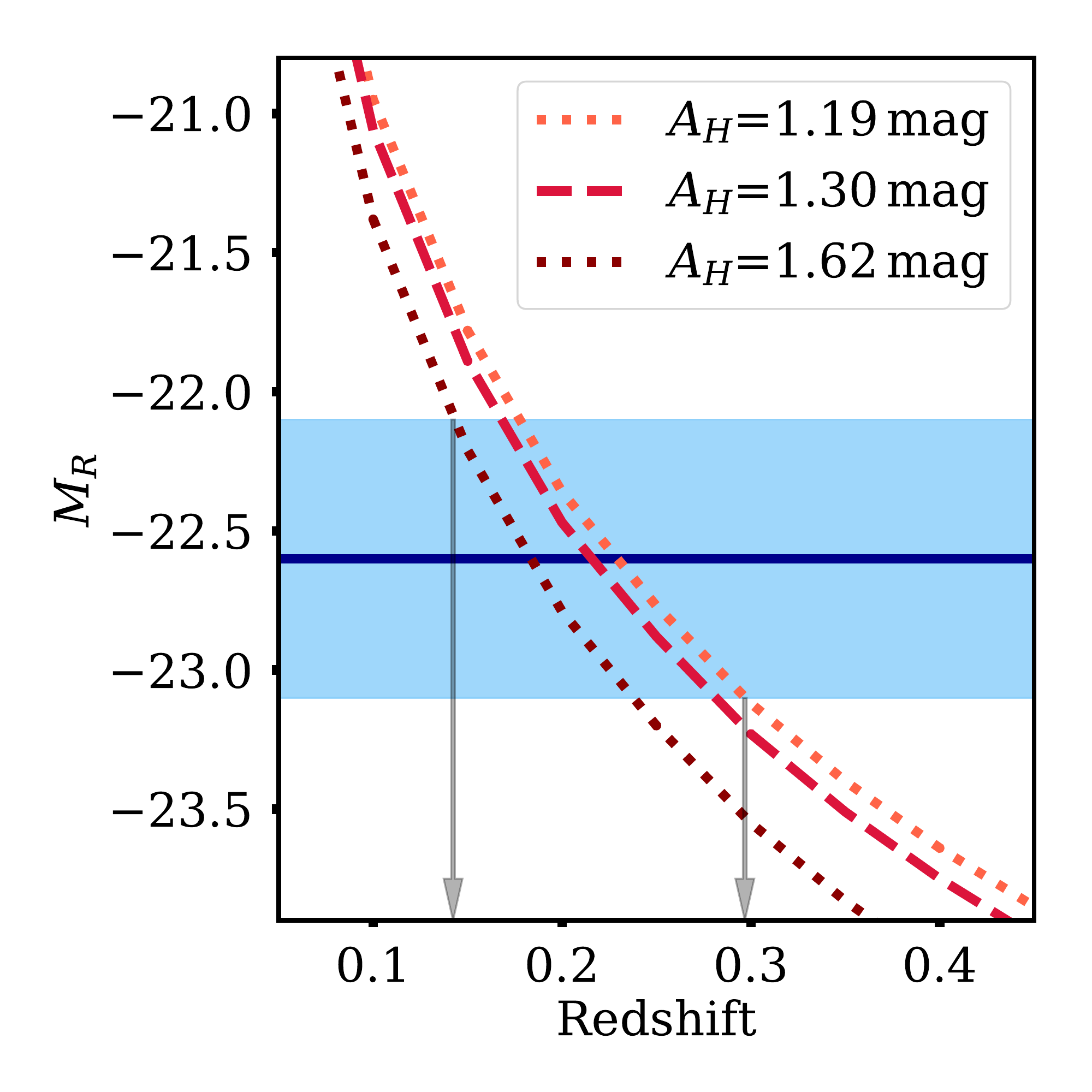}
\caption
{
Expected rest--frame $R$--band absolute magnitude of the host galaxy HESS~J1943+213 as a function of redshift (see Section~\ref{sec:host} for details).
Dashed and dotted lines shows the effect of different level of dust extinction.
The average M$_R$ of BL~Lac host galaxies derived by \citet{Sbarufatti2005b} is shown as an horizontal blue line, with the $\pm1\sigma$ regime highlighted in light blue.
The redshift of HESS~J1943+213 is thus constrained in the range 0.14$\lesssim$$z$$\lesssim$0.30 (vertical grey arrows).
}
\label{fig:redshift}
\end{center}
\end{figure}

Table~\ref{tab:fit} summarises the results of our fit.
The nuclear (unresolved) emission is $H$$_{\rm PSF}$=18.10\,mag, while the host galaxy appears to be brighter with $H$$_{\rm Host}$=16.15\,mag.
The derived values for the effective radius ($R_e$=1.13\,arcsec) and for the S{\'e}rsic index ($n$=2.20) are in contrast with the 2\farcs0$\lesssim$$R_e$$\lesssim$2\farcs5 and $n$$\sim$8 estimated by \citealt{Peter2014}.
These where derived from the analysis of a $Ks$--band image collected with the wide field camera OMEGA2000 on the 3.5\,m Calar Alto telescope.
The smaller values we recover for both $R_e$ and $n$ are consistent with an independent analysis performed with the {\textsc Imfit} package \citep[][]{Erwin2015}, where instead of masking it, we also fit for the source (star) projected close to the blazar.
We argue that the discrepancies between the Peter et al. and our study are due to the much higher spatial resolution delivered by ARGOS\footnote{Seeing during the observations of HESS~J1943+213 at Calar Alto was between 1\farcs1 and 1\farcs6.}.
In addition, Peter et al. model the blazar emission with a single S{\'e}rsic profile (i.e., without removing the PSF).
An underestimate of the contribution from the nuclear emission may explain the high S{\'e}rsic index observed in the OMEGA2000 image.

Given the magnitude of the host galaxy, we can now estimate the redshift of HESS~J1943+213 using the typical absolute magnitude of BL~Lac host galaxies as a standard candle.
Indeed, \citet{Sbarufatti2005b} showed that, at $z$$<$0.7, the distribution of the rest--frame $R$--band absolute magnitude of BL~Lac host galaxies is almost Gaussian, with an average of $\langle$M$_{\rm R}$$\rangle$=-22.6 and standard deviation $\sigma_{\rm M_{\rm R}}$=0.5.
To translate, as a function of redshift, the observed $H$--band apparent magnitude into an $R$--band absolute magnitude we considered the \citet[][]{Mannucci2001} elliptical galaxy template and a passive evolution of the stellar population \citep[][]{Bressan1998}.
This allowed us to find at which redshift the observed magnitude of the host galaxy match $\langle$M$_{\rm R}$$\rangle$=-22.6.
Assuming A$^{\rm c}_{\rm H}$=1.30\,mag we derive a redshift $z$=0.21 (varying from $z$=0.17 to $z$=0.28 for $\langle$M$_{R}$$\rangle$-$\sigma_{\rm M_{R}}$ and 
$\langle$M$_{R}$$\rangle$+$\sigma_{\rm M_{R}}$, respectively).
At this redshift, the physical effective radius of the host galaxy is $R_e$=3.9\,kpc.
We also took into account the effects of the uncertainties in the Galactic extinction presented in Section~\ref{sec:dr}.
These led to slightly different results, ranging from $z$=0.23 for $A^{c}_{H}$=1.19\,mag to $z$=0.18 for $A^{c}_{H}$=1.62\,mag (see Figure~\ref{fig:redshift}).

\section{Summary and Conclusions}

We obtained deep LBT/LUCI\,1 $H$--band imaging of the blazar HESS~J1943+213 detected by {\it H.E.S.S.} in the VHE demain.
The superb spatial resolution (FWHM=0\farcs26) delivered by the new adaptive optics system, ARGOS, allowed us to precisely separate the unresolved nuclear component from the extended host galaxy emission.

The host galaxy of HESS~J1943+213 appears round (with axis ratio of $\sim$0.92), with a S{\'e}rsic index $n$$\sim$2.2, and with $H$--band magnitude of $H_{\rm Host}$$\sim$16.15\,mag.
Assuming the host galaxy as a standard candel, we locate HESS~J1943+213 at $z$$\sim$0.21, or, more conservatively, in the redshift range 0.14$<$$z$$<$0.30 (considering variation of the typical luminosity of BL~Lac host galaxies and uncertainties in the Galactic extinction).
This range is consistent with the limits set by comparing measured spectra in high energy (HE, $E$$<$100\,GeV) and very high energy (VHE, $E>100$\,GeV) ranges.
Assuming any difference between the two was a consequence of absorption of the VHE $\gamma$--rays by the EBL, \citet{Peter2014} were able to set the upper limit to $z$$<$0.45, with the most likely value of $z$=0.22.
Our result agrees very well with their estimate, and our upper limit of $z$$<$0.30 is well below the one set by \citet{Peter2014}.
The spectra measured by {\it VERITAS} \citep{Shahinyan2017} in the 2014-2015 period are slightly harder, although consistent with {\it H.E.S.S.} measurements.
Therefore, the extrapolated and EBL--attenuated {\it Fermi}--LAT spectrum should also fit {\it VERITAS} spectral points.
Despite the fact that we set the most stringent limits on redshift up to date, the uncertainties are still rather high, allowing for the measured VHE spectra to be a result of combination of intrinsic spectral features and the EBL--absorption.

In addition, we detect a source located only 0\farcs6 West from the BL~Lac.
Although it is very likely a foreground star, if it is at the same redshift as HESS~J1943+213, this corresponds to a physical distance of $\sim$2.1\,kpc.
This may be suggestive for a rich environment, as commonly observed around other BL~Lac host galaxies.

\section*{Acknowledgements}

EPF acknowledges funding through the ERC grant `Cosmic Dawn'.
EPF is grateful to T.\,A.~Gutcke for introducing and providing support for an efficient use of \textsc{Python} and of \textsc{Jupyter Notebook} for analysing and plotting the data.
This research made use of \textsc{Astropy}, a community--developed core \textsc{Python} package for Astronomy \citep{Astropy2013}, of \textsc{APLpy}\footnote{\texttt{http://aplpy.github.io/}}, an open-source plotting package for \textsc{Python} based on \textsc{Matplotlib} \citep{Hunter2007}, and of \textsc{IRAF}\footnote{\textsc{IRAF} \citep{Tody1986, Tody1993}, is distributed by the National Optical Astronomy Observatories, which are operated by the Association of Universities for Research in Astronomy, Inc., under cooperative agreement with the National Science Foundation.}.
Based on observations collected at the Large Binocular Telescope (LBT).
The LBT is an international collaboration among institutions in the United States, Italy and Germany. LBT Corporation partners are: The University of Arizona on behalf of the Arizona Board of Regents; Istituto Nazionale di Astrofisica, Italy; LBT Beteiligungsgesellschaft, Germany, representing the Max-Planck Society, The Leibniz Institute for Astrophysics Potsdam, and Heidelberg University; The Ohio State University, and The Research Corporation, on behalf of The University of Notre Dame, University of Minnesota and University of Virginia.







\bsp	
\label{lastpage}
\end{document}